\documentstyle[12pt,epsf,amssymb]{article}

\def\hybrid{\topmargin -20pt    \oddsidemargin 0pt
        \headheight 0pt \headsep 0pt
        \textwidth 6.25in       
        \textheight 9.5in       
        \marginparwidth .875in
        \parskip 5pt plus 1pt   \jot = 1.5ex}

\hybrid

\def\baselinestretch{1.2}

\catcode`\@=11

\def\marginnote#1{}
\newcount\hour
\newcount\minute
\newtoks\amorpm
\hour=\time\divide\hour by60
\minute=\time{\multiply\hour by60 \global\advance\minute by-\hour}
\edef\standardtime{{\ifnum\hour<12 \global\amorpm={am}%
        \else\global\amorpm={pm}\advance\hour by-12 \fi
        \ifnum\hour=0 \hour=12 \fi
        \number\hour:\ifnum\minute<10 0\fi\number\minute\the\amorpm}}
\edef\militarytime{\number\hour:\ifnum\minute<10 0\fi\number\minute}

\def\draftlabel#1{{\@bsphack\if@filesw {\let\thepage\relax
   \xdef\@gtempa{\write\@auxout{\string
      \newlabel{#1}{{\@currentlabel}{\thepage}}}}}\@gtempa
   \if@nobreak \ifvmode\nobreak\fi\fi\fi\@esphack}
        \gdef\@eqnlabel{#1}}
\def\@eqnlabel{}
\def\@vacuum{}
\def\draftmarginnote#1{\marginpar{\raggedright\scriptsize\tt#1}}

\def\draft{\oddsidemargin -.5truein
        \def\@oddfoot{\sl preliminary draft \hfil
        \rm\thepage\hfil\sl\today\quad\militarytime}
        \let\@evenfoot\@oddfoot \overfullrule 3pt
        \let\label=\draftlabel
        \let\marginnote=\draftmarginnote
   \def\@eqnnum{(\theequation)\rlap{\kern\marginparsep\tt\@eqnlabel}%
\global\let\@eqnlabel\@vacuum}  }

\def\preprint{\twocolumn\sloppy\flushbottom\parindent 2em
        \leftmargini 2em\leftmarginv .5em\leftmarginvi .5em
        \oddsidemargin -.5in    \evensidemargin -.5in
        \columnsep .4in \footheight 0pt
        \textwidth 10.in        \topmargin  -.4in
        \headheight 12pt \topskip .4in
        \textheight 6.9in \footskip 0pt
        \def\@oddhead{\thepage\hfil\addtocounter{page}{1}\thepage}
        \let\@evenhead\@oddhead \def\@oddfoot{} \def\@evenfoot{} }

\def\numberbysection{\@addtoreset{equation}{section}
        \def\theequation{\thesection.\arabic{equation}}}

\def\underline#1{\relax\ifmmode\@@underline#1\else
        $\@@underline{\hbox{#1}}$\relax\fi}

\def\titlepage{\@restonecolfalse\if@twocolumn\@restonecoltrue\onecolumn
     \else \newpage \fi \thispagestyle{empty}\c@page\z@
        \def\thefootnote{\fnsymbol{footnote}} }

\def\endtitlepage{\if@restonecol\twocolumn \else \newpage \fi
        \def\thefootnote{\arabic{footnote}}
        \setcounter{footnote}{0}}  

\catcode`@=12
\relax

\def\figcap{\section*{Figure Captions\markboth
        {FIGURECAPTIONS}{FIGURECAPTIONS}}\list
        {Figure \arabic{enumi}:\hfill}{\settowidth\labelwidth{Figure
999:}
        \leftmargin\labelwidth
        \advance\leftmargin\labelsep\usecounter{enumi}}}
 \relax
\def\tablecap{\section*{Table Captions\markboth
        {TABLECAPTIONS}{TABLECAPTIONS}}\list
        {Table \arabic{enumi}:\hfill}{\settowidth\labelwidth{Table
999:}
        \leftmargin\labelwidth
        \advance\leftmargin\labelsep\usecounter{enumi}}}
 \relax
\def\reflist{\section*{References\markboth
        {REFLIST}{REFLIST}}\list
        {[\arabic{enumi}]\hfill}{\settowidth\labelwidth{[999]}
        \leftmargin\labelwidth
        \advance\leftmargin\labelsep\usecounter{enumi}}}
 \relax
\makeatletter
\newcounter{pubctr}
\def\publist{\@ifnextchar[{\@publist}{\@@publist}}
\def\@publist[#1]{\list
        {[\arabic{pubctr}]\hfill}{\settowidth\labelwidth{[999]}
        \leftmargin\labelwidth
        \advance\leftmargin\labelsep
        \@nmbrlisttrue\def\@listctr{pubctr}
        \setcounter{pubctr}{#1}\addtocounter{pubctr}{-1}}}
\def\@@publist{\list
        {[\arabic{pubctr}]\hfill}{\settowidth\labelwidth{[999]}
        \leftmargin\labelwidth
        \advance\leftmargin\labelsep
        \@nmbrlisttrue\def\@listctr{pubctr}}}
 \relax
\makeatother
\newskip\humongous \humongous=0pt plus 1000pt minus 1000pt

\newif\ifdtup

\relax

\def\be{\begin{equation}}
\def\ee{\end{equation}}
\def\ba{\begin{eqnarray}}
\def\ea{\end{eqnarray}}

\def\r{\rho}
\def\a{\alpha}

\def\G{\Gamma}
\def\d{\delta}

\def\m{\mu}

\def\Om{\Omega}
\def\l{\lambda}

\def\S{\Sigma}

\def\tm{\tilde{m}}

\def\no{\noindent}

\def\qq{\qquad}

\def\IR{\relax{\rm I\kern-.18em R}}

\def \ha {{1\over 2}}

\def \ov {\over}

\def\IR{\relax{\rm I\kern-.18em R}}
\def\inv{^{\raise.15ex\hbox{${\scriptscriptstyle -}$}\kern-.05em 1}}

\def\tL{{\tilde L}}

\begin{document}

\renewcommand{\theequation}{\arabic{equation}}

\newcommand{\beq}{\begin{equation}}
\newcommand{\eeq}[1]{\label{#1}\end{equation}}
\newcommand{\ber}{\begin{eqnarray}}
\newcommand{\eer}[1]{\label{#1}\end{eqnarray}}
\newcommand{\eqn}[1]{(\ref{#1})}
\begin{titlepage}
\begin{center}

\hfill CERN-TH/99-148\\
\hfill hep--ph/9905417\\

\vskip .8in

{\large \bf Deviations from the $1/r^2$ Newton law  due to  extra 
 dimensions }

\vskip 0.6in

{\bf A. Kehagias}\phantom{x} and\phantom{x} {\bf K. Sfetsos}
\vskip 0.1in
{\em Theory Division, CERN\\
     CH-1211 Geneva 23, Switzerland\\
{\tt kehagias,sfetsos@mail.cern.ch}}\\
\vskip .2in

\end{center}

\vskip .6in

\centerline{\bf Abstract }

\no
We systematically 
examine corrections to the gravitational inverse square law, which are due to 
compactified extra dimensions. 
We find the induced Yukawa-type potentials for which we calculate
the strength $\alpha$ and range. In general the range of the Yukawa correction 
is given by the wavelength of the lightest Kaluza--Klein state and its
strength, relative to the standard gravitational potential,
by the corresponding degeneracy. 
In particular, when $n$ extra dimensions are compactified on an $n$-torus,
we find that the strength of the potential is $\alpha=2n$,
whereas the compactification on an $n$-sphere gives $\a= n+1$.
For Calabi--Yau compactifications the strength can be at most $\a=20$.

\vskip 0,2cm
\no

\vskip 3cm
\noindent
CERN-TH/99-148\\
May 1999\\
\end{titlepage}
\vfill
\eject

\def\baselinestretch{1.2}
\baselineskip 16 pt
\noindent

\def\tT{{\tilde T}}
\def\tg{{\tilde g}}
\def\tL{{\tilde L}}


\section{Introduction and discussion }

Recently, deviations of the inverse square law for  gravity have   
received a lot of attention \cite{ADD}--\cite{L}. 
In general, these deviations are 
parametrized by two parameters, $\alpha $ and $\lambda$, corresponding 
to the strength, with respect to the $1/r^2$ law, and the range. Specifically,
the form of the potential is (for experimental aspects, see, for instance,
\cite{L} and references therein) 
\be
V(r)=-\frac{G_4 M}{r}\left(1+\alpha e^{-r/\lambda}\right)\ , 
\label{pll}
\ee
where $G_4$ is the four-dimensional Newton constant and $M$ is the mass.
This form of the potential is expected to be valid for $r\gg \l$
and, in general, there will be more terms correcting the $1/r$ potential, 
which are nevertheless subdominant, as we will also see.

There are experimental bounds on the possible values of $\a$ and $\l$, 
which are
represented in an $\a$--$\l$ diagram \cite{ADD,L}, at the end of the paper. 
The value of $\l$ is restricted to be 
at most of order $1$ mm, leaving the possibility of new forces in the 
submillimeter regime \cite{ADD,AHDD}.  
On the other hand, as depicted in the figure,
there are theoretical models that can give 
different values of the strength $\a$. For example, it was argued, 
in \cite{ADD}, that a Scherk--Schwarz 
supersymmetry-breaking mechanism at $1$ TeV 
gives rise to a scalar radius modulus and a potential of the form 
\eqn{pll} with $\alpha\sim 4/7$, whereas a mechanism involving the dilaton 
predicts $\a\sim 44$ \cite{TV,ADD}. 

In this letter we systematically 
examine corrections to the $1/r$ gravitational potential
due to extra dimensions. 
We consider Einstein gravity in $n+4$ dimensions, where $n$ is the number 
of extra dimensions, and we find the Newtonian limit of the theory. Then
we compactify the internal $n$ dimensions in order to obtain 
the four-dimensional effective
gravitational potential; to leading order, this
is of the form \eqn{pll} with $\l$ proportional
to the inverse mass of the lightest Kaluza--Klein (KK)
state and $\a$ equal to its  degeneracy. We explicitly derive this result
in section 2 for a general compactification manifold. 
In particular, for the case of an $n$-dimensional torus, 
we find, since the number of extra dimensions can be $n=2,3,\ldots,7$,
that the strength can take the values $\a=4,6,\ldots,14$.\footnote{There 
cannot be only one extra dimension ($n=1$)
because the deviation from Newtonian
gravity would then have been over astronomical distances \cite{AHDD}.
The upper value in the number of extra dimensions, $n=7$, corresponds
to a compactification from the highest-dimensional consistent theory,
which is eleven-dimensional supergravity.} We also discuss the cases of
sphere compactification, where the strength can take the values $\a=3,4,\dots,
8$ and Calabi--Yau (CY) compactification, where we argue that $\a\leq 20$.

Note that there exist other possibilities such as torsion, massive 
gravitinos, Brans--Dicke scalars etc., which are expected
 to produce similar 
corrections to the Newton law and
it would be interesting to study them.

\section{Gravitational potential and extra dimensions}

In this section we consider the corrections 
to the gravitational potential
due to extra dimensions. We will first  
calculate the potential for the case of compactification on 
an $n$-dimensional torus and then on other spaces, including on the 
$n$-dimensional sphere and CY manifolds.

\subsection{Toroidal compactification}

We assume that the space-time is ($n+4$)-dimensional, where the $n$ extra 
dimensions $x_i$, $i=1,2,\dots, n$, are compactified on circles, each 
with radius $R_i$. The Newtonian limit of a ($n+4$)-dimensional 
black hole will give the gravitational potential of a massive object. Since, 
to our knowledge, higher-dimensional black-hole solutions with some dimensions
compactified are not known, we will examine the Newtonian limit of 
higher-dimensional gravity and 
we will impose compactification on the solution.
Presumably, the result can also be obtained in the Newtonian limit of a,
yet unknown, higher-dimensional black hole.\footnote{A 
periodic-black-hole solution in four space-time dimensions has 
been constructed in \cite{KN}.}
 
The gravitational  potential of a massive object with mass $M$ at  
a distance $r_{n}= \left(r^2 + x_1^2+x_2^2+\dots +x_n^2\right)^{1/2}$, 
where  $r^2=x^2+y^2+z^2$ is the three-dimensional radial distance, 
satisfies the $(n+3)$-dimensional Laplace equation, and it is given by   
\be
V_{n+4} = -\sum_{{\bf m}\in {\bf Z}} {G_{n+4} M \ov\Big(
r^2 + \sum_{i=1}^n(x_i- 2 \pi R_i m_i)^2\Big)^{(n+1)/2}}\ . 
\label{pott} 
\ee
Here, $G_{n+4}$ is the Newton constant in $n+4$ dimensions and ${\bf m}=(m_1,
m_2,\dots ,m_n)$ is a vector in a $n$-dimensional lattice. 
This potential satisfies the
appropriate boundary conditions, namely, it vanishes at spatial infinity and 
it is periodic in the extra $n$ dimensions since it is invariant under the 
shifts $x_i\rightarrow x_i+2\pi R_i$. 
For very large $R_i$'s only the term with ${\bf m}= {\bf 0}$ survives in the 
sum and we recover the familiar Newton law in $n+4$ dimensions:
\be
V_{n+4} \simeq -  
{ G_{n+4} M \ov r_{n}^{n+1}} \ .
\label{ndhj}
\ee
On the other hand, if the $R_i$'s are small, we may approximate the sum by an 
integral as
\be
V_{n+4} \simeq -{G_{n+4}M\ov \S_n}
 \int d^n {\bf x} {1\ov 
(r^2 + {\bf x}^2)^{(n+1)/2}}=- {\Om_n G_{n+4} M \ov 2 \S_n} 
\ {1\ov r}\ , 
\label{apw}
\ee
where the volume of the $n$-dimensional torus $\S_n$ and that 
of the $n$-dimensional
unit sphere $\Om_n$ are given by 
\be
\S_n =(2\pi)^n \prod_{i=1}^n R_i\ , \qq 
\Om_n= {2 \pi^{n+1\ov 2}\ov \Gamma\left({n+1\ov 2}\right)}\ .
\label{vnom}
\ee
By comparing \eqn{apw} with the potential in four space-time dimensions
$V_4=-{G_4 M\ov r}$, the four-dimensional Newton constant is identified as
\be
G_4 = {\Om_n G_{n+4} \ov 2 \S_n} \ .
\label{nnnne}
\ee
This relation, and the observed value of the four-dimensional Planck scale
$G_4^{1/2}\sim 10^{-33}$ cm,
leads to a unification of the Planck-scale in $n+4$ space-time dimensions
(with $n\geq 2$)
with the electroweak interactions scale $1$ TeV  (or $10^{-20}$ m),
provided that the typical compactification radius of the circles 
is $R\sim 1$ mm or smaller \cite{AHDD}, thus
realizing previous proposals for large 
internal dimensions \cite{Witten}.
In turn, that suggests a novel resolution of the hierarchy problem 
\cite{AHDD,Bachas}.

In order to discuss deviations from Newtonian gravity, we must compute 
the first corrections to \eqn{apw}.
This is done by  Poisson resuming \eqn{pott} and we 
obtain\footnote{The two integrals 
below are computed using the formulae 8.411(8) and 
6.565(3) of \cite{tipologio}.}
\ba
&&V_{n+4}  =  -{G_{n+4} M \ov \S_n} \sum_{{\bf m}\in {\bf Z}} 
\int d^n {\bf x} {e^{-i {\bf  \tm}\cdot {\bf x}}\ov  
\Big(r^2 + \sum_{i=1}^n(x_i- 2 \pi R_i m_i)^2 \Big)^{{n+1\ov 2}}  }
\nonumber \\
&& =  - 
{2\Om_{n-2}G_{n+4} M\ov \S_n} \sum_{{\bf m}\in {\bf Z}} 
e^{-i {\bf  \tm}\cdot {\bf x}}
\int_0^\infty d\r 
{\r^{n-1}\ov (r^2+\r^2)^{(n+1)/2}} \int_0^1 dx  \cos(|{\bf \tm}| \r x)
(1-x^2)^{n-3\ov 2} 
\nonumber\\
&& = 
-{\Om_{n-2} {2^{n/2} \sqrt{\pi} \G\left({n-1\ov 2}\right)} \ov 2 \S_n} 
\ G_{n+4} M \sum_{{\bf m}\in {\bf Z}} 
{e^{-i {\bf  \tm}\cdot {\bf x}}\ov  |{\bf \tm}|^{n/2-1}}  
\int_0^\infty d\r \
{\r^{n/2} J_{{n\ov 2}-1}(|{\bf \tm}| \r) \ov (r^2+\r^2)^{(n+1)/2} }\ ,
\label{jashh}
\ea
where $J_{{n\ov 2}-1}$ is the Bessel function of order ${n\ov 2}-1$ and
${\bf \tm} =\left({m_1\ov R_1},\dots, {m_n\ov R_n}\right)$. Note that 
$|{\bf \tm}|=\left({m_1^2\ov R_1^2} +{m_2^2\ov R_2^2} +\dots +
{m_n^2\ov R_n^2} \right)^{1/2}$ are the masses of the KK-modes. 
After performing the last integral in \eqn{jashh} we find 
\be 
V_{n+4} =  - {G_4 M\ov r } \ \sum_{{\bf m}\in {\bf Z}} e^{-r |{\bf \tm}|}\
e^{-i {\bf  \tm}\cdot {\bf x}}\ ,
\label{jash}
\ee
where $G_4$ is defined in \eqn{nnnne}.
Next we omit the 
internal space  dependence since all point particles in  
the four-dimensional space-time can be taken to  have ${\bf x}={\bf 0}$.
Hence, the four-dimensional gravitational potential, in the presence of $n$
extra dimensions compactified on a $n$-dimensional torus, is given 
by\footnote{This result has been also obtained 
in \cite{arkd} and by E. Floratos and G. Leontaris (to appear).
We thank R. Rattazzi for bringing \cite{arkd} into our attention.}
\be 
V_{4} =  - {G_4 M\ov r } \ \sum_{{\bf m}\in {\bf Z}} e^{-r |{\bf \tm}|}\ .
\label{jashe}
\ee
It is clear from the above expression that the Newton $1/r$ potential
results from
the term in the sum with ${\bf m}={\bf 0}$. The first correction to it comes 
from the lightest KK states.
Thus, we find that the gravitational potential is approximately of the type 
\eqn{pll}, namely
\be
V_{4} \simeq -{G_4 M\ov r } \left( 1+ 2 n_0 e^{- r/R_0} \right)\ ,
\label{rtps}
\ee
where $1/R_0$ is the lightest KK mass and $2n_0$ is its degeneracy and
 $n_0$ is the number of equal radii,  
i.e. $R_1=\ldots=R_{n_0}\equiv R_0$. 
Thus we see that the strength equals the degeneracy of the lightest 
KK state and the range is its wavelength.

\subsection{Compactification on other manifolds}

Let us consider a space-time of the form ${\rm Minkowski} \times M^n$, 
where $M^n$ is an $n$-dimensional 
compact manifold; let $\{\Psi_{\bf m}\}$  be a set of functions
in $M^n$ obeying the orthogonality condition
\be
\int_{M^n} \Psi_{\bf n}({\bf x}) \Psi^*_{\bf m}({\bf x}) = \d_{{\bf n},{\bf m}} \ ,
\label{oort}
\ee
as well as the completeness relation 
\be
\sum_{{\bf m}} \Psi_{\bf m}({\bf x}) \Psi^*_{\bf m}({\bf x'}) = \d^{(n)}({\bf x},
{\bf x'})\ .
\label{oort1}
\ee
The functions  $\{\Psi_{\bf n}\}$ are eigenfunctions of the $n$-dimensional
Laplace operator with  eigenvalues $\m^2_{\bf m}$
\be
\nabla_n^2  \Psi_{\bf m} = - \m_{{\bf m}}^2  \Psi_{\bf m} \ .
\label{mssa}
\ee
In the Newtonian limit, the 
gravitational potential $V_{n+4}$ satisfies the Poisson equation  
in $n+3$ spatial dimensions:
\be \nabla_{n+3}^2 V_{n+4} =  (n+1) \Om_{n+2} G_{n+4} M \d^{(n+3)}({\bf x})\ ,
\label{jaaa}
\ee
which is solved by $V_{n+4} =- {G_{n+4} M\ov r_n^{n+1}}$. For $n$ compact
dimensions, we may  expand $V_{n+4}$ in terms of the complete basis of
eigenfunctions of the Laplace operator on $M^n$,
$\{\Psi_{{\bf m}}\}$ as
\be 
V_{n+4}= \sum_{{\bf m}} \Phi_{{\bf m}}(r) \Psi_{{\bf m}}({\bf x})\ .
\label{jan}
\ee
Then the $\Phi_{{\bf m}}$'s obey 
\be
\nabla_3^2 \Phi_{{\bf m}} - \m^2_{{\bf m}} \Phi_{{\bf m}} = (n+1) \Om_{n+2}
\Psi^*_{{\bf m}}({\bf 0}) {G_{n+4} M} \d^{(3)}({\bf x})\ ,
\label{jdsh}
\ee
with solution 
\be 
\Phi_{{\bf m}}(r) = - {\Om_n G_{n+4} M \Psi^*_{{\bf m}}({\bf 0})\ov 2 }\ 
{1\ov r}\ e^{-\m_{\bf m} r}\ ,
\label{jdhsf}
\ee
so that \eqn{jash} changes to
\be
V_{n+4} =- {\Om_n G_{n+4} M\ov 2 r }
\sum_{{\bf m}}   
\Psi^*_{\bf m}({\bf 0}) \Psi_{\bf m}({\bf x})\ e^{-\m_{\bf m} r }\ .
\label{rtps1}
\ee
As explained before we may omit the internal space dependence and set 
${\bf x}={\bf 0}$ in the above formula.
Then we may further simplify it by realizing that the sum over ${\bf m}$ 
is over all possible allowed 
irreducible representations of the symmetry group of the
compact manifold $M^n$ and then, for each such representation, over
all representatives. However, the eigenvalue of the Laplace operator 
$\m_{\bf m}$ depends only on the representation and not on the particular
representative that was used to compute it in \eqn{mssa}. Then, from
\eqn{rtps1} we obtain the four-dimensional gravitational potential in the 
presence of $n$ extra dimensions compactified on a general manifold $M^n$, as
\be
V_{4}= -{G_4 M\ov r} \sum_{\bf m_{\rm ir}} d_{\bf m_{\rm ir}}
e^{-\m_{\bf m_{\rm ir}} r}\ ,
\label{djs}
\ee
where we sum up over all possible irreducible representations 
${\bf m_{\rm ir}}$, and $d_{\bf m_{\rm ir}}$ denotes the corresponding 
degeneracy. The four-dimensional Newton constant $G_4$
is defined as in \eqn{nnnne},
where $\S_n$ is the volume of the compact manifold $M^n$. In passing from 
\eqn{rtps1} to \eqn{djs} we have also 
used the group theoretical result that the 
sum of $|\Psi_{{\bf m_{\rm ir}}}|^2$ 
over all representatives of a given irreducible 
representation equals $d_{\bf m_{\rm ir}}/\S_n$.
Using this general formula we see that, to leading order for large $r$,
the gravitational potential is of the form \eqn{pll}, with range inversely
proportional to the mass of the lightest KK state and strength equal to its
degeneracy.  The general result (\ref{djs}) reduces to (\ref{jash}) for the 
case of compactification on an n-torus. In that case, the symmetry group 
is abelian and $d_{\bf m_{\rm ir}}=1$. 

\subsubsection{Compactification on spheres}

Let us illustrate these by first considering the $n$-dimensional sphere of
radius $R$ as our 
compactification manifold. A general KK state has mass and degeneracy 
given by \cite{vN} 
\ba
&&\m_m= {\sqrt{m(m+n-1)}\ov R}\ ,\qq m =0,1,\dots \ ,
\nonumber \\
&& {d}_{m} = {(2 m +n-1) (m+n-2)!\ov (n-1)! m!}\ .
\label{made}
\ea
Then \eqn{djs} takes the form 
\be
V_{4}= -{G_{4} M\ov r} \sum_{m=0}^\infty {d}_m \ 
e^{-\m_m r} \ ,
\label{jqw}  
\ee
where, the Newton constant is 
\be
G_4 = {G_{n+4} \ov 2 R^n }\ .
\label{nne1}
\ee
This potential is approximately, for large r, 
\be
V_4 \simeq - {G_4 M \ov r} \left(1 + (n+1) e^{-\sqrt{n}\ r/R} \right)\ .
\label{trkl}
\ee
Note that the range of the induced Yukawa potential is given by the 
mass of the lightest KK state, whereas its strength is its degeneracy, which
is $n+1$, namely the dimension of the vector representation of $SO(n+1)$. 
It is instructive to compare the
strengths of  the Yukawa-type correction for compactifications on
the $n$-sphere and  on the $n$-dimensional 
torus. For the $n$-sphere, the strength of the Yukawa-type
correction is $\a=3,4,\dots , 8$, whereas for the $n$-torus 
$\a=4,\dots , 14$. Hence, the topology of the compactification manifold of
the extra dimensions seems 
to be hard to detect experimentally, since the strengths are comparable.  

\subsubsection{Compactification on Calabi--Yau manifolds}

Theories with  $N=1$ supersymmetry in four-dimensions are obtained by 
CY compactifications in string theory. 
The CY manifolds are Ricci-flat K\"ahler manifolds 
with no continuous isometries and the explicit
metric for them is not known. Hence, it is not possible to even attempt 
solving the eigenvalue equation \eqn{mssa}. However, we may compute the 
degeneracy of the eigenstates using well known group theoretical results.

Typically, for CY manifolds with a (discrete) global symmetry group there
exists a symmetry factor containing products of the permutation group 
$S_n$ and of the cyclic group $Z_n^\ell$($\equiv Z_n\times\dots\times 
Z_n$).\footnote{A way 
to construct manifolds with $SU(3)$ holonomy is to start
with the $N$-dimensional complex projective space ($CP_N$) and place enough
constraints that reduce its complex dimensions to three.
For example, for $CP_4$ we put $\sum_{i=1}^4 z_i^5 =0$, for 
$CP_3\times CP_3$ we put $\sum_{i=1}^3 z_i^3 =0$, $\sum_{i=1}^3 w_i^3 =0$
and $\sum_{i=1}^3 z_i w_i =0$.}
The irreducible representations of $S_n$ are labelled my a set of $n$
non-negative integers $\{m_i\}$, subject to the constraint \cite{hamer}
\be
m_1\ge m_2\ge \dots \ge m_n\ge 0\ ,\qq  m_1 + m_2 +\dots +  m_n = n \ .
\label{conr}
\ee
The dimensionality of an irreducible representation is given by 
\be
d_{\bf m} = {n!\ov h_1!h_2!\dots h_n! } \prod_{i<j}^n (h_i-h_j)\ ,\qq
h_i \equiv m_i +n-i\ .
\label{dimrw}
\ee
The lowest dimensional massless state corresponds to the solution of \eqn{conr}
with $m_1=n$ and $m_2=m_3=\dots =m_n=0$. It is a singlet under both $S_n$ and  
$Z_n^\ell$ consistent with the fact that the Hodge number 
$h^{0,0}=1$ for all CY spaces. 
The first massive state is in the lowest non-trivial 
representation of $S_n$, which  corresponds to
$m_1=m_2=\dots =m_n=1$, and it is one-dimensional as can be seen from
\eqn{dimrw}. This state is  degenerate in $Z_n^\ell$ so that its degeneracy 
is at least $n \ell$.
In the cases of CY manifolds with no symmetries at all, the  lowest
bound for the degeneracy of the first massive state is of cource (we bear 
in mind accidental degeneracies)
\be
d_{\rm lower} =  {\cal O}(1)\ .
\label{lwps}
\ee
For a ``maximally symmetric'' isotropic quintic, the symmetry group is 
isomorphic to the semi-direct product of $S_5$ and $Z_5^4$. 
Using this model, we obtain the upper bound for the degeneracy of the
first massive state as
\be
d_{\rm upper} =  20\ .
\label{lwp1s}
\ee
Hence,  
the strength of the Yukawa-type correction to the inverse square law
associated with the CY compactification 
can be at most $\a=20$ which is a bit larger, but nevertheless comparable, to  
the values $\a=12$ and $\a=7$ for the torus
$T^6$ and sphere $S^6$, respectively.

Similar  corrections to the Newton law are expected to arise from other 
sources such as torsion, massive 
gravitinos, Brans--Dicke scalars etc.,  which however have not discussed here. 
We have collected our results in fig. 1.

\begin{figure}[htb]
\epsfxsize=4in
\bigskip
\centerline{\epsffile{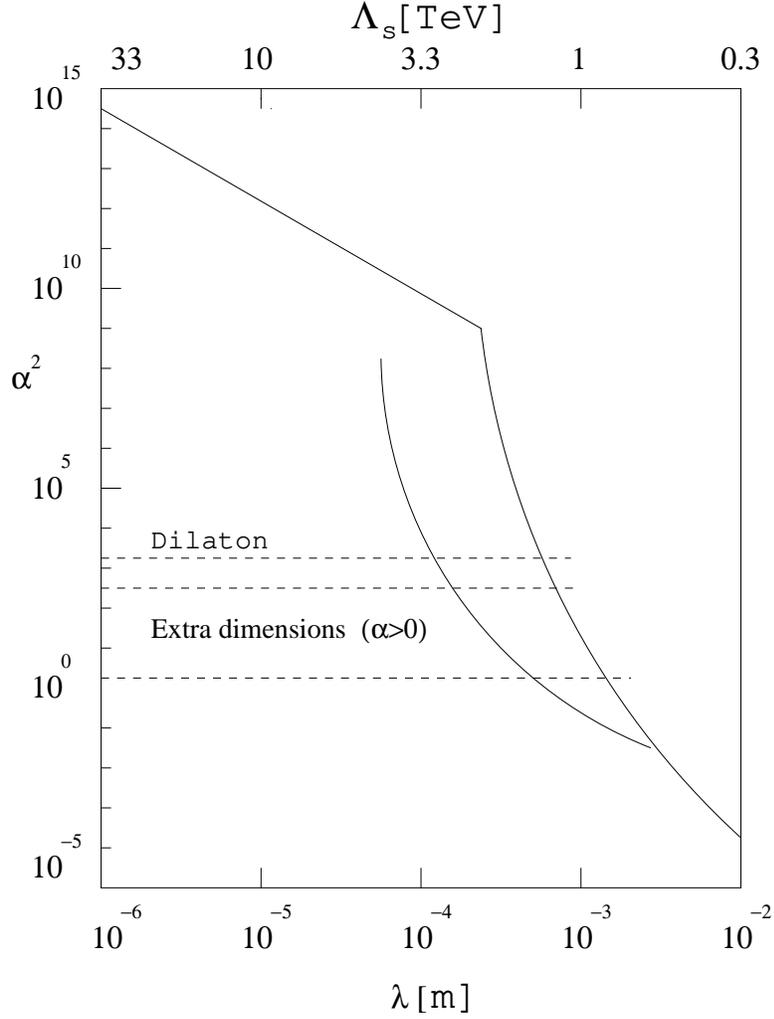}}
\caption{Plot of the $\a$--$\l$ diagram where the experimental bounds 
are indicated by solid lines according to \cite{ADD,L}. Our predictions 
for the strength in the case of extra dimensions range between $\a^2=
1,\dots, 9,\dots,
196,\dots, 400$ 
as indicated by the two $(---)$ lines.  We also 
indicated the value $\a^2\sim 2000$, which corresponds 
to the dilaton conntribution. }
\bigskip
\label{fig1}
\end{figure}


\bigskip\bigskip

\centerline{\bf Acknowledgements}
We thank R. Rattazzi for a very useful correspondence.



\end{document}